\documentclass[pra,aps,floatfix,twocolumn,superscriptaddress,amsmath,amssymb]{revtex4-2}
\usepackage{graphicx}
\usepackage{color}
\usepackage{bm}
\usepackage{hyperref}
\hypersetup{colorlinks=true,citecolor=blue,linkcolor=blue,urlcolor=blue}
\usepackage{CJKutf8}
\usepackage{braket}
\usepackage{mleftright}
\allowdisplaybreaks
\usepackage{graphicx}
\usepackage[normalem]{ulem}

\begin{document}

\title{%
Surface criticality in the mixed-field Ising model with sign-inverted next-nearest-neighbor interaction
}

\author{Yuki Nakamura}
\affiliation{%
Graduate School of Human and Environmental Studies, Kyoto University, Kyoto 606-8501, Japan}
\affiliation{%
Department of Physics, Kindai University, Higashi-Osaka, Osaka 577-8502, Japan}

\author{Ryui Kaneko}
\email{ryuikaneko@sophia.ac.jp}
\affiliation{%
Department of Physics, Kindai University, Higashi-Osaka, Osaka 577-8502, Japan}
\affiliation{%
Physics Division, Sophia University, Chiyoda, Tokyo 102-8554, Japan}

\author{Ippei Danshita}
\email{danshita@phys.kindai.ac.jp}
\affiliation{%
Department of Physics, Kindai University, Higashi-Osaka, Osaka 577-8502, Japan}

\date{\today}

\begin{abstract}
Rydberg atoms in an optical tweezer array have been used as a quantum simulator of the spin-$1/2$ antiferromagnetic Ising model with longitudinal and transverse fields. 
We suggest how to implement the next-nearest-neighbor (NNN) interaction whose sign is opposite to that of the nearest neighbor one in the Rydberg atom systems. 
We show that this can be achieved by weakly coupling one Rydberg state with another Rydberg state.  
We further study the surface criticality associated with the first-order quantum phase transition between the antiferromagnetic and paramagnetic phases, which emerges due to the sign-inverted NNN interaction.
From the microscopic model, we derive a Ginzburg-Landau (GL) equation, which describes static and dynamic properties of the antiferromagnetic order parameter near the transition. Using both analytical GL theory and numerical method based on a mean-field theory, we calculate the order parameter in the proximity of a boundary of the system in order to show that the healing length of the order parameter logarithmically diverges, signaling the surface criticality.

\end{abstract}

\maketitle

\section{Introduction}
Rapid technological progress in preparing and manipulating Rydberg atoms in an optical tweezer array has led to successful applications of the systems as quantum simulators of quantum Ising models~\cite{Browaeys,Morgado,Wu}. Specifically, recent experiments have reported observations of various magnetically ordered phases, such as antiferromagnetic phases breaking ${\mathbb Z}_n$ symmetry ($n=2, 3,$ and $4$) in 1D chain~\cite{Bernien}, phases with checkerboard, striated, and star patterns in a square lattice~\cite{Ebadi}, and three-sublattice ordered phases with $1/3$ and $2/3$ fillings in a triangular lattice~\cite{Scholl}. Such developments have triggered renewed theoretical interest of quantum phases and quantum phase transition in Ising-type models~\cite{Lienhard,Rader,Kaneko}. 

Since the spin-spin interaction of the Ising model simulated by Rydberg atoms corresponds to van der Waals interaction between two atoms in a Rydberg state, it decays steeply with the distance $r$ as $\propto r^{-6}$. While a situation where the next-nearest-neighbor (NNN) interaction is comparable to the nearest-neighbor (NN) one can be implemented with use of Rydberg-dressed states~\cite{Henkel,Zeiher}, it does not allow for changing the sign of the interactions. Previous theoretical studies have shown that the NNN interaction whose sign is opposite to that of the NN one is a key to accessing some interesting phases and transitions, including the floating phase in a 1D chain~\cite{Beccaria} and quantum tricriticality in a square lattice~\cite{Kato}. In this sense, for extending the applicability of Rydberg-atom quantum simulators, it is useful to enable the sign inversion of the NNN interaction.

In this paper, we propose that the Ising model with sign-inverted NNN interaction can be realized by using a Rydberg state coupled weakly to another Rydberg state. As an interesting phenomenon that can be studied in the proposed system, we investigate the surface criticality~\cite{Lipowsky,Lipowsky87,Danshita} emerging near the first-order quantum phase transition between the antiferromagnetic and paramagnetic phases in a square lattice. We derive a Ginzburg-Landau (GL) equation describing the motion of the antiferromagnetic order parameter and relate the coefficients in the GL equation to the parameters in the microscopic model. By calculating the order parameter in the proximity of the boundary with use of both analytical GL theory and numerical method based on a mean-field theory, we show the logarithmic divergence of the healing length of the order parameter, which is a characteristic feature of the surface criticality.

The remainder of the paper is organized as follows. 
In Sec.~\ref{sec:NNN}, we explain how to incorporate sign-inverted NNN interaction in the Ising model with use of a Rydberg state coupled with another Rydberg state.
In Sec.~\ref{sec:surface}, we analyze the surface criticality near the first-order quantum phase transition of the Ising model with the sign-inverted NNN interaction. We derive the GL equation for gaining analytical insights and examine the critical behavior by numerically solving the mean-field equation.
In Sec.~\ref{sec:summary}, we summarize the results. We set $\hbar = 1$ throughout the paper.

\section{How to create sign-inverted next-nearest neighbor interaction}
\label{sec:NNN}
\begin{figure}
    \includegraphics[scale=0.5]{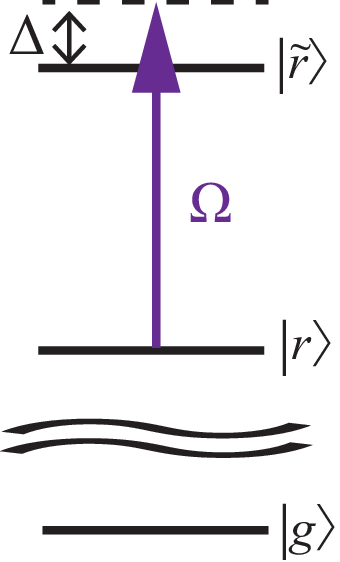}
    \caption{Schematic illustration of a Rydberg state $\ket{r}$ that is weakly coupled to another Rydberg state $\ket{\tilde{r}}$.}
\label{fig:rdr}
\end{figure}

In this section, we aim to show that the Ising model with the sign-inverted NNN interaction can be created by using a Rydberg state coupled slightly with another Rydberg state. We consider the system consisting of $M$ neutral atoms spatially aligned on a square lattice with an optical tweezer array. We assume that one of the electronically excited states with a large principal quantum number $n$, namely Rydberg states, is coupled with the ground state via a two-photon Raman process. We represent the ground state and the Rydberg state of the $j$th atom as $|g_j\rangle$ and $|r_j\rangle$, respectively. We focus on the interaction between the $j$th atom and the $l$th atom ($j\neq l$) whose positions are denoted by ${\bm x}_j$ and ${\bm x}_l$, respectively. We assume that the lattice spacing $a$ is sufficiently large so that the interaction is negligible if either of the two atoms is in the ground state. We also assume that the Rydberg state has even parity, as
there is no direct dipole-dipole interaction between the two Rydberg atoms. Hence, the dominant interaction between the two Rydberg atoms is the van der Waals interaction given by
\begin{eqnarray}
V_{rr}(|{\bm x}_{jl}|)=\frac{C_6}{|\bm{x}_{jl}|^6},
\end{eqnarray}
where $C_6$ is the interaction coefficient.
In addition, as illustrated in Fig.~\ref{fig:rdr}, we consider a situation where the Rydberg state $|r_j\rangle$ is slightly coupled with another Rydberg state with even parity, denoted by $|\tilde{r}_j\rangle$. The interaction between two atoms in $|r_j\tilde{r}_l\rangle$ state and that between two atoms in $|\tilde{r}_j\tilde{r}_l\rangle$ state are of van der Waals type given by
\begin{eqnarray}
V_{r\tilde{r}}(|{\bm x}_{jl}|)=\frac{\tilde{C}_6}{|\bm{x}_{jl}|^6},
\\
V_{\tilde{r}\tilde{r}}(|{\bm x}_{jl}|)=\frac{D_6}{|\bm{x}_{jl}|^6},
\end{eqnarray}
where $\tilde{C}_6$ and $D_6$ are the interaction coefficients for $V_{r\tilde{r}}(|{\bm x}_{jl}|)$ and $V_{\tilde{r}\tilde{r}}(|{\bm x}_{jl}|)$, respectively.
When the Hilbert space of a single atom is spanned with only $|r_j\rangle$ and $|\tilde{r}_j\rangle$, the Hamiltonian of the two atoms can be written as
\begin{eqnarray}
\hat{H}_{\rm two} = \hat{H}_{\rm two}^{(0)} + \hat{V},
 \label{eq:like dressed hamiltonian}
\end{eqnarray}
where
\begin{equation}
\begin{split}
   \hat{H}_{\rm two}^{(0)}&=-\Delta\left(\hat{n}_j+\hat{n}_l\right)+V_{\tilde{r}\tilde{r}}\left(|\bm{x}_{jl}|\right)\hat{n}_j\hat{n}_l\\
    &\quad+V_{rr}\left(|\bm{x}_{jl}|\right)\left(1-\hat{n}_j\right)\left(1-\hat{n}_l\right)\\
    &\quad+V_{r\tilde{r}}\left(|\bm{x}_{jl}|\right)\left(\left(1-\hat{n}_j\right)\hat{n}_l+\hat{n}_j\left(1-\hat{n}_l\right)\right)
\end{split}
\end{equation}
and
\begin{equation}
    \hat{V}=\frac{\Omega}{2}\left(\hat{X}_j+\hat{X}_l\right). \label{eq:laser coupling}
\end{equation}
Here $\Delta$ and $\Omega$ denote the detuning and Rabi frequency of the coupling between $|r_j\rangle$ and $|\tilde{r}_j\rangle$. The operators $\hat{n}_j$ and $\hat{X}_j$ are defined as
\begin{eqnarray}
\hat{n}_j &=& |\tilde{r}_j\rangle\langle\tilde{r}_j|,
\\
\hat{X}_j &=& |r_j\rangle\langle\tilde{r}_j| + |\tilde{r}_j\rangle\langle r_j|.
\end{eqnarray}

Assuming $\Delta \gg \Omega$, we regard $\hat{V}$ as a perturbation to the unperturbed part of the Hamiltonian $\hat{H}_{\rm two}^{(0)}$. The energies of the unperturbed states, $\ket{r_j r_l}$\,,\,$\ket{r_j\tilde{r}_l}$, $\ket{\tilde{r}_j r_l}$,  and $\ket{\tilde{r}_j\tilde{r}_l}$ are
\begin{eqnarray}
    E^{\left(0\right)}_{rr} &=& V_{rr}\left(|{\bm x}_{jl}|\right),\\
    E^{\left(0\right)}_{\tilde{r}r}  = E^{\left(0\right)}_{r\tilde{r}} &=& -\Delta+V_{r\tilde{r}}\left(|{\bm x}_{jl}|\right),\\
    E^{\left(0\right)}_{\tilde{r}\tilde{r}} &=& -2\Delta+V_{\tilde{r}\tilde{r}}\left(|{\bm x}_{jl}|\right). \label{eq:nonpertubation}
\end{eqnarray}
When the unperturbed state is $|\Phi^{(0)}\rangle = |r_jr_l\rangle$, the first-order perturbed state and the second-order perturbation energy read
\begin{eqnarray}
 |\Phi\rangle &=&
 |r_jr_l\rangle  
 \nonumber \\
 &&\!\!\!\!\! + \frac{\Omega}{\Delta - V_{r\tilde{r}}(|{\bm x}_{jl}|) + V_{rr}(|{\bm x}_{jl}|)}\left( |r_j\tilde{r}_l\rangle \! +\! |\tilde{r}_jr_l\rangle  \right),
\end{eqnarray}
and
\begin{equation}
    E_{rr}^{\left(2\right)} = \frac{\Omega^2}{2\Delta} + \frac{F_6}{\left|{\bm x}_{jl}\right|^6+\tilde{R}^6}, \label{like 2nd pertubation}
\end{equation}
where $F_6=\frac{\Omega^2}{2\Delta^2}\left(\tilde{C_6}-C_6\right)$ and $\tilde{R}=\left(-\frac{\tilde{C}_6-C_6}{\Delta}\right)^{\frac{1}{6}}$. Hence, the energy $E_{rr}$ up to the second-order perturbation is given by
\begin{equation}
    E_{rr}\simeq E_{rr}^{(0)}+E_{rr}^{(2)}= \frac{C_6}{|{\bm x}_{jl}|^6}+\frac{\Omega^2}{2\Delta}+\frac{F_6}{|{\bm x}_{jl}|^6+\tilde{R}^6}. 
    \label{eq:effective-potential}
\end{equation}
This means that the interaction between the two atoms both in the Rydberg-dressed Rydberg (RDR) state is given by $V^{\rm RDR}(|{\bm x}_{jl}|) = E_{rr}$.
Combining the interaction with the local magnetic field terms~\cite{Browaeys}, one can write the many-body Hamiltonian as a form of the spin-1/2 Ising model
\begin{equation}
    \hat{H}_{\rm IM}=\frac{1}{2}\sum_{ j\neq l}J_{jl}\hat{\sigma}_j^z\hat{\sigma}_l^z-\sum_j \left(\frac{\delta_j}{2} - U_j\right)\hat{\sigma}_j^z-\Gamma\sum_j\hat{\sigma}_j^x, \label{eq:Hamiltonian}
\end{equation}
where $\hat{\sigma}_j^z$ and $\hat{\sigma}_j^x$ are the $z$ and $x$ components of the Pauli matrices at site $j$. We rewrite the interaction as $J_{jl}=4V^{\rm RDR}(|{\bm x}_{jl}|)$ for the brevity of the notation. $\delta_j$ and $2\Gamma$ correspond to the detuning and frequency of the Rabi coupling between $|g_j\rangle$ and $|r_j\rangle$. The collective contribution $U_{j}=\sum_{l\neq j}J_{jl}$ arises as a consequence of the mapping from the original atomic representation to the effective spin-1/2 representation. While $U_j$ is inhomogeneous near the boundaries of the system, such inhomogeneity can be 
canceled 
by controlling the detuning $\delta_j$ such that $\frac{\delta_j}{2}-U_j=H$ is homogeneous. In the spin-1/2 representation, $\ket{\downarrow}$ corresponds to the atomic ground state while $\ket{\uparrow}$ does to the RDR state.

\begin{figure}
 \includegraphics[width=\linewidth]{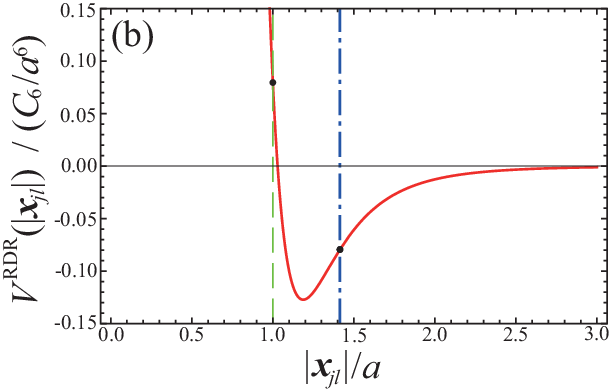}
 \caption{Effective interaction potential $V^{\rm RDR}(|{\bm x}_{jl}|)$ between the two atoms both in the RDR state as a function of the interatomic distance ${\bm x}_{jl}$, where we set $F_6/C_6=-1.841$. }
 \label{fig:VRDR}
\end{figure}

One can see from Eq.~(\ref{eq:effective-potential}) that if $F_6$ has a sign opposite to $C_6$ and $|F_6|>|C_6|$, $V^{\rm RDR}(|{\bm x}_{jl}|)$ changes its sign at 
\begin{eqnarray}
|{\bm x}_{jl}|= \left(\frac{|C_6|}{|F_6|-|C_6|}\right)^{\frac{1}{6}}\tilde{R}.
\end{eqnarray}
For example,  in Fig.~\ref{fig:VRDR}, we show $V^{\rm RDR}(|{\bm x}_{jl}|)$ as a function of $|{\bm x}_{jl}|$ for $\tilde{R}/a = 1$, $C_6>0$, and $F_6/C_6 = -1.841$. There $V^{\rm RDR}(|{\bm x}_{jl}|=a) = -V^{\rm RDR}(|{\bm x}_{jl}|=\sqrt{2}a)>0$, meaning that the NNN interaction has a sign opposite to the NN one on a square lattice and their magnitudes are equal. Moreover, the interaction for $|{\bm x}_{jl}|>\sqrt{2}a$ decays abruptly as $\sim |{\bm x}_{jl}|^{-6}$ so that it is negligible if $\max(|\Gamma|,|H|) \gg |V^{\rm RDR}(|{\bm x}_{jl}|=2a)|$.
Under these assumptions, the Ising model reads
\begin{equation}
    \hat{H}_{\rm IM}=J_1\sum_{\langle j,l\rangle}\hat{\sigma}_j^z\hat{\sigma}_l^z-J_2\sum_{\langle\langle j,l\rangle\rangle}\hat{\sigma}_j^z\hat{\sigma}_l^z-H\sum_j\hat{\sigma}_j^z-\Gamma\sum_j\hat{\sigma}_j^x 
    \label{eq:HamiltonianNNN}
\end{equation}
where $J_1 =4V^{\rm RDR}(|{\bm x}_{jl}|=a)$ and $J_2 =-4V^{\rm RDR}(|{\bm x}_{jl}|=\sqrt{2}a)$
denote the NN and NNN spin-spin interactions.

Let us take $^{87}$Rb atoms, which have been indeed used in several experiments regarding Rydberg atom arrays~\cite{Bernien,Ebadi,Scholl,Lienhard}, as an example for more specifically explaining correspondence between the above mentioned protocol and actual experiments. The atomic states involved in the protocol, $|g\rangle$, $|r\rangle$, and $|\tilde{r} \rangle$ can be chosen as $|5S_{1/2},F=2,m_F=-2\rangle$, $|nS_{1/2},J=1/2,m_J=-1/2\rangle$, and $|\tilde{n}D_{3/2},J=3/2\rangle$, where $\tilde{n}\geq n$ and $n\gtrsim 50$. The coupling between $|g\rangle$ and $|r\rangle$ can be made via a two-photon Raman process with an intermediate state $|e\rangle$. For instance, in Ref.~\cite{Bernien}, $|70S_{1/2},J=1/2,m_J=-1/2\rangle$ and $|6P_{3/2},F=3,m_F=-3\rangle$ has been used as $|r\rangle$ and $|e\rangle$ and the transition between $|g\rangle$ and $|e\rangle$ and that between $|e\rangle$ and $|r\rangle$ are driven by lasers with wave-length 420 nm and 1013 nm, respectively. The $|\tilde{r}\rangle$ state has to be chosen in a way such that the van der Waals interaction between $|r_j\rangle$ and $|\tilde{r}_l\rangle$ satisfies that $\tilde{C}_6<0$ and $|\tilde{C}_6|\gg C_6$, which imply that $F_6<0$ and $C_6 \sim |F_6|$. The coupling between $|r\rangle$ and $|\tilde{r} \rangle$ can be made via a two-photon Raman process or a single-photon process using microwave.

In the existing method using Rydberg-dressed states (see, e.g., Refs.~\cite{Henkel,Zeiher}), an electronically ground state of atoms, whose van der Waals interaction is negligibly small, is slightly coupled with a Rydberg state so that the interaction between two dressed atoms includes only a fourth-order perturbation term of the coupling. This does not allow for changing the sign of the atom-atom interaction. On the contrary, in our proposed method, a Rydberg state $|r\rangle$ of atoms, whose van der Waals interaction is rather large, is slightly coupled with another Rydberg state $|\tilde{r}\rangle$ so that the interaction between two dressed atoms includes a second order perturbation term of the coupling as well as the bare van der Waals interaction between two atoms both in state $|r\rangle$. A main advantage is that one can change the sign of the two-body interaction, which has been exploited for preparing the sign-inverted NNN interaction in the proposed protocol. 

We conclude this section by discussing one of the main practical challenges in the above proposed experimental setups, i.e., a search of Rydberg states suited for the second Rydberg state $|\tilde{r}\rangle$. Under the assumption that $C_6>0$, while the van der Waals interaction between $|r_j\rangle$ and $|\tilde{r}_l\rangle$ has to satisfy that $\tilde{C}_6<0$ and $|\tilde{C}_6|\gg C_6$ such that the effective interaction satisfies $F_6<0$ and $C_6 \sim |F_6|$, it has not been known that which Rydberg state indeed satisfies those conditions. A potential solution for this challenge is to seek such a Rydberg state by experimentally determining the van der Waals interaction constant $\tilde{C}_6$. For example, direct measurement of the van der Waals interaction coefficient has been reported in Ref.~\cite{Beguin}, where the shift of the resonance energy of the $|r_jr_l\rangle$ state from the $|r_jg_l\rangle$ state due to the Rydberg blockade was measured by analyzing the Rabi oscillation between the two states.

\section{Surface criticality near the first-order quantum phase transition}
\label{sec:surface}
\begin{figure}
 \includegraphics[width=\linewidth]{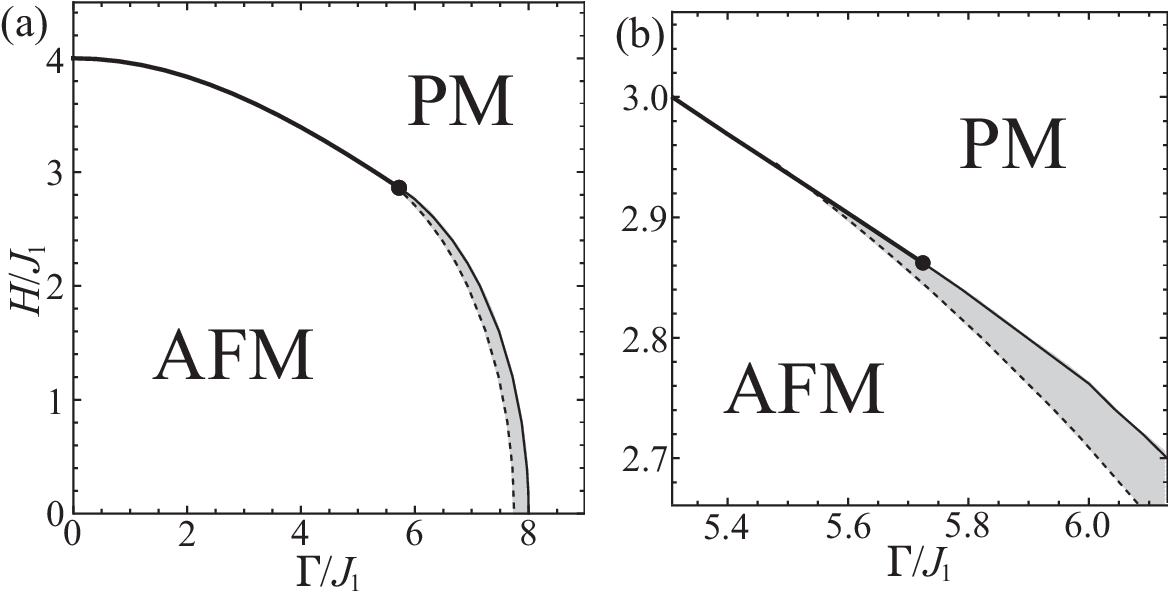}
 \caption{(a) Ground-state phase diagram of the Hamiltonian (\ref{eq:HamiltonianNNN}) in the $(\Gamma/J_1,H/J_1)$-plane for $J_1=J_2$ and $J_1>0$. The thick (thin) solid line represents the phase boundary of the first (second)-order quantum phase transition between the antiferromagnetic (AFM) and paramagnetic (PM) phases, which has been obtained in Ref.~\cite{Kato}. The filled circle marks the quantum tricritical point (QTCP). The dashed line represents the contour of $\phi_0=0.25$, where $\phi_0$ is the value of the order parameter in the bulk, so that the gray shaded region roughly indicates the antiferromagnetic (AFM) region where the GL theory is valid. (b) Magnified view of (a) near the QTCP.}
 \label{fig:phaseD}
\end{figure}
We briefly review the ground-state phase diagram of the mixed-field Ising model with the sign-inverted NNN interaction of Eq.~(\ref{eq:HamiltonianNNN}), which has been previously obtained in Ref.~\cite{Kato} by means of a mean-field theory.
In Fig.~\ref{fig:phaseD}, we show the phase diagram in the $(\Gamma/J_1,H/J_1)$-plane, where $J_2/J_1=1$ and $J_1>0$. There are two distinct phases in the phase diagram. One is the antiferromagnetic (AFM) phase, which is favored for small magnetic fields compared to the spin-spin interaction. The AFM phase has a two-sublattice structure. The other is the paramagnetic (PM) phase for large magnetic fields. When $H\ll \Gamma$, the transition between the two phases is of second order. When $H\gtrsim \Gamma$, it is of first order. The point at which the transition changes from second- to first-order one on the transition line is the quantum tricritical point (QTCP). Within a mean-field theory, the QTCP is located at 
\begin{eqnarray}
(\Gamma/J_1,H/J_1) = \frac{32}{5\sqrt{5}}\times\left(2,1\right)\simeq 2.862\times\left(2,1\right).
\end{eqnarray}
Notice that a more quantitative phase diagram has been obtained also in Ref.~\cite{Kato} with use of the quantum Monte Carlo method. Here we show the mean-field phase diagram simply because we use the mean-field theory throughout the present paper.

In the remainder of this section, we analyze the surface criticality that emerges near the first-order transition. 
In the context of phase transitions, critical phenomena generally stem from the divergence of a length scale. One of the most well-known examples is a second-order phase transition where the correlation length of a certain correlation function diverges at a phase transition point, which is called a critical point. Near the critical point, several observables, including thermodynamic quantities, exhibit universal critical behavior. On the other hand, in the case of first-order transitions, the correlation length in a bulk does not diverge at the transition point such that critical phenomena do not emerge in a bulk. However, if one focuses on the vicinity of the surfaces (boundaries or edges) of a system, there is a length scale that logarithmically diverges at the first-order transition point. Supposing that the order parameter vanishes at the surface, this length scale quantifies the distance from the surface at which the order parameter revives to be comparable to the bulk value so that we call it the healing length. Associated with the logarithmic divergence of the healing length, several quantities, such as the order parameter amplitude near the surface and the surface free energy, exhibit critical behavior, which is termed as the surface criticality~\cite{Lipowsky,Lipowsky87}.

In the case of classical phase transitions, the surface criticality has been experimentally observed in the solid-liquid interface of several substances, such as Pb~\cite{Frenken} and water~\cite{Bluhm,Limmer}. On the contrary, to the best of our knowledge, there has been no experimental observation of the surface criticality in quantum phase transitions while it was theoretically proposed in Ref.~\cite{Danshita} that the surface criticality can occur in two-component ultracold Bose gases in optical lattices.

The presence of the QTCP is helpful for studying the surface criticality in the sense that one can widely control the surface critical region. While it is known that the mixed field Ising model with van der Waals-type spin-spin interaction on a square lattice, which has been already realized in experiments of Ref.~\cite{Ebadi}, exhibits first-order quantum phase transitions between ordered and disordered phases, there is no QTCP in the ground-state phase diagram of that model~\cite{Samajdar,Kalinowski}.

Throughout the present paper, we use a mean-field theory, which ignores the effects of quantum fluctuations. As shown in Ref.~\cite{Lipowsky83}, the upper critical dimension of the surface criticality in classical phase transitions is $d=3$, where $d$ denotes the spatial dimension of a classical system. Moreover, numerical Monte Carlo simulations of classical Ising models with sign-inverted NN and NNN interactions on fcc lattice \cite{Gompper} showed that the mean-field theory captures qualitative features of the surface criticality at $d=3$, such as the logarithmic dependence of the healing length and the critical exponent. Given that the dynamical exponent is unity in the quantum phase transition between the antiferromagnetic and paramagnetic phases, the upper critical dimension for our quantum surface criticality is $D=2$, where $D$ denotes the spatial dimension of a quantum system. Hence, we expect that our mean-field approximation is also valid for qualitatively describing surface criticality in a two-dimensional quantum system. 

Nevertheless, there are 
certain
limitations of the mean-field approach. A main limitation is the lack of quantitativeness. For instance, the quantum tricritical point in the ground-state phase diagram of our model is given as $(\Gamma/J_1,H/J_1)=(5.684,2.842)$ within the mean-field approximation while the quantum Monte Carlo simulations, which is exact within statistical errors, gives $(\Gamma/J_1,H/J_1)=(4.10(5),3.260(2))$~\cite{Kato}. For more quantitative analyses of the quantum surface criticality, quantum Monte Carlo method might be useful. However, it is rather challenging to carry out such numerical analyses using quantum Monte Carlo method for a system that is large enough to address the surface criticality, say $L>O(100)$, where $L$ is the system size in one direction in a unit of the lattice constant. Hence, we leave such more quantitative numerical analyses for future work.

\subsection{Ginzburg-Landau theory}
We assume that the system is in a parameter region indicated in Fig.~\ref{fig:phaseD} as the gray-shaded region, where the antiferromagnetic order parameter
\begin{eqnarray}
\psi({\bm x}_j,t)=(2z_1)^{-1}\sum_{\langle l\rangle_j} \left(\langle \hat{\sigma}^z_{j}\rangle - \langle \hat{\sigma}^z_{l}\rangle\right)
\end{eqnarray}
at the position ${\bm x}_j$ and the time $t$ is sufficiently small so that the GL theory is approximately valid. Here, $\langle l\rangle_j$ means the NN sites to site $j$ and $z_1$ denotes the coordination number of NN bonds, which is 4 in the case of the square lattice considered in the present study.
There, the motion of the antiferromagnetic order parameter 
near the first-order phase transition is described by the GL equation
\begin{equation}
    -w\frac{\partial^2\psi}{\partial t^2}=\left(-C\nabla^2+s+u\psi^2+v\psi^4\right)\psi.
    \label{eq:GL}
\end{equation}
The expression of the coefficients $s$, $u$, and $v$ in terms of the Ising parameters have been obtained for zero temperature and $z_1J_1=z_2J_2$ in Ref.~\cite{Kato} as
\begin{eqnarray*}
    s&=&J_+\left(1-\frac{\Gamma^2J_+}{B^3}\right),\\
    u&=&\frac{\left(\Gamma^2-4H^2\right)\Gamma^2J_+^4}{2B^7},\\
    v&=&\frac{3\left(12\Gamma^2H^2-8H^4-\Gamma^4\right)\Gamma^2J_+^6}{8B^{11}},\\
    J_+ &=& z_1J_1 + z_2 J_2,
   \\
    B&=&\sqrt{\Gamma^2+H^2},
\end{eqnarray*}
where $z_2$ is the coordination number of NNN bonds ($z_2=4$ for the square lattice).

We next relate the remaining constants $w$ and $C$ to the Ising parameters from the spin-wave dispersion relation. For this purpose, we assume that the order parameter takes the form of 
\begin{eqnarray}
\psi({\bm x},t) = \phi({\bm x}) + \delta \psi({\bm x},t),
\label{eq:smallfluc}
\end{eqnarray}
where $\phi({\bm x})$ is a stationary order parameter and $\delta \psi({\bm x},t)$ is a small fluctuation of the order parameter from $\phi({\bm x})$.
Substituting Eq.~(\ref{eq:smallfluc}) into Eq.~(\ref{eq:GL}) and neglecting second and higher order terms with respect to $\delta \psi$, one obtains the stationary GL equation,
\begin{eqnarray}
\left(-C\nabla^2+s+u\phi^2+v\phi^4\right)\phi=0,
\label{eq:stationaryGL}
\end{eqnarray}
and the linearized equation,
\begin{eqnarray}
-w\frac{\partial^2}{\partial t^2}\delta\psi=\left(-C\nabla^2+s+3u\phi^2+5v\phi^4\right)\delta\psi.
\label{eq:linearizedGL}
\end{eqnarray}
Assuming a paramagnetic solution $\phi=0$ of Eq.~(\ref{eq:stationaryGL}) and substituting a plane-wave solution $\delta \psi = e^{i({\bm p}\cdot {\bm x}-\omega t)}$ into Eq.~(\ref{eq:linearizedGL}), one obtains the dispersion relation of the small fluctuations,
\begin{equation}
    \omega=\sqrt{\frac{C}{w}p^2+\frac{s}{w}}.\label{eq:bunsankankei_GL}
\end{equation}

We also calculate the dispersion relation of the small fluctuations in the paramagnetic phase by applying a mean-field approximation~\cite{Danshita-10} directly to the original model of Eq.~(\ref{eq:HamiltonianNNN}) in order to compare its long-wavelength limit to Eq.~(\ref{eq:bunsankankei_GL}).
We start with the Heisenberg's equations of motion for $\hat{\sigma}_j^z$ and $\hat{\sigma}_j^- = \left(\hat{\sigma}_j^x -i\hat{\sigma}_j^y\right)/2$,
\begin{gather}
    \frac{d\hat{\sigma}_j^z}{dt}=-2\Gamma\hat{\sigma}_j^y, \label{eq:Heisenberg1}
    \\
\begin{split}
    i\frac{d\hat{\sigma}_j^-}{dt}&=2J_1\sum_{\langle l\rangle_j}\hat{\sigma}_j^-\hat{\sigma}_l^z-2J_2\sum_{\langle\langle l \rangle\rangle_j}\hat{\sigma}_j^-\hat{\sigma}_l^z
    \\
    &\quad-2H\hat{\sigma}^{-}_j+\Gamma \hat{\sigma}_j^z, \label{eq:Heisenberg2}
\end{split}
\end{gather}
where $\langle\langle l\rangle\rangle_j$ means the NNN sites to site $j$.
We assume that the many-body wave function is approximated as a direct product of local spin \sout{coherent} states,
\begin{eqnarray}
\!\!\!\!\!\!\!\!|\Phi_{\rm MF}\rangle\! =\! \bigotimes_j \left(e^{-i\frac{\varphi_j}{2}}\cos\frac{\theta_j}{2}\ket{\uparrow}_j + e^{i\frac{\varphi_j}{2}}\sin\frac{\theta_j}{2}\ket{\downarrow}_j \right),
\end{eqnarray}
where $\theta_j$ and $\varphi_j$ denote the elevation and azimuthal angles of the spin direction at site $j$.
Replacing $\hat{\sigma}_j^z$ and $\hat{\sigma}_j^{-}$ in Eqs.~(\ref{eq:Heisenberg1}) and (\ref{eq:Heisenberg2}) with their mean fields $\langle\hat{\sigma}_j^z\rangle=\cos{\theta_j}$ and $\langle\hat{\sigma}_j^-\rangle=\frac{1}{2}e^{-i\varphi_j}\sin{\theta_j}$, one obtains the mean-field equations of motion,
\begin{equation}
\begin{split}
    \frac{d\varphi_j}{dt}\sin{\theta_i}&=2J_1\sin{\theta_j}\sum_{\langle l\rangle_j}\cos{\theta_l}-2J_2\sin{\theta_j}\sum_{\langle\langle l\rangle\rangle_j}\cos{\theta_l}   \label{eq:MF1}
\\
    &\quad-2H\sin{\theta_j}+2\Gamma\cos{\theta_j\cos{\varphi_j}},
\end{split}
\end{equation}
\begin{equation}
    \frac{d\theta_j}{dt}=2\Gamma\sin{\varphi_j}. \label{eq:MF2}
\end{equation}
Since we are interested in the dispersion relation of the spin wave excitations, we assume that the solutions of Eqs.~(\ref{eq:MF1}) and (\ref{eq:MF2}) take the following forms,
\begin{eqnarray}
\theta_j(t) &=& \bar{\theta}_j +\delta \theta_j(t),
\label{eq:deltheta}
\\
\varphi_j(t) &=& \bar{\varphi}_j + \delta \varphi_j(t),
\label{eq:delphi}
\end{eqnarray}
where $\bar{\theta}_j$ and $\bar{\varphi}_j$ are the stationary solutions while $\delta \theta_j(t)$ and $\delta \varphi_j(t)$ are small fluctuations from the stationary solutions.
Substituting Eqs.~(\ref{eq:deltheta}) and (\ref{eq:delphi}) into Eqs.~(\ref{eq:MF1}) and (\ref{eq:MF2}), and neglecting second and higher order terms with respect to the fluctuations, one obtains the equations for the stationary solutions,
\begin{equation}
\begin{split}
    0=&\,\,2J_1\sin{\Bar{\theta}_j}\sum_{\langle l\rangle_j}\cos{\Bar{\theta}_l}-2J_2\sin{\Bar{\theta}_j}\sum_{\langle\langle l\rangle\rangle_j}\cos{\Bar{\theta}_l}
    \\
    &-2H\sin{\Bar{\theta}_j}+2\Gamma\cos{\Bar{\theta}_j}\cos{\Bar{\varphi}_j},
    \label{eq:stationaryMF1}
\end{split}
\end{equation}
\begin{equation}
    0=\Gamma\sin{\Bar{\varphi}_j},
    \label{eq:stationaryMF2}
\end{equation}
and the linearized equations of motion,
\begin{equation}
\begin{split}
    &\frac{d}{dt}\delta\varphi_j=2J_1\left(-\sum_{\langle l\rangle_j}\sin{\bar{\theta}_l}\delta\theta_l+\frac{\cos{\Bar{\theta}_j}}{\sin{\bar{\theta}_j}}\sum_{\langle l\rangle_j}\cos{\Bar{\theta}_l}\delta\theta_j\right)\\
    &\quad-2J_2\left(-\sum_{\langle\langle l\rangle\rangle_j}\sin{\bar{\theta}_l}\delta\theta_l+\frac{\cos{\Bar{\theta}_j}}{\sin{\bar{\theta}_j}}\sum_{\langle\langle l\rangle\rangle_j}\cos{\Bar{\theta}_l}\delta\theta_j\right)\\
    &\quad-2H\frac{\cos{\bar{\theta}_j}}{\sin{\bar{\theta}_j}}\delta\theta_j-2\Gamma\left(\cos{\Bar{\varphi}_j}\delta\theta_j+\frac{\cos{\Bar{\theta}_j}}{\sin{\bar{\theta}_j}}\sin{\Bar{\varphi}_j}\delta\varphi_j\right) ,
    \label{eq:linearizedMF1}
\end{split}    
\end{equation}
\begin{equation}
    \frac{d}{dt}\delta\theta_j=2\Gamma\cos{\Bar{\varphi}_j}\delta\varphi_j. \label{eq:linearizedMF2}
\end{equation}
We assume that the system is in the paramagnetic phase, where $\bar{\varphi}_j = 0$. Moreover, we recall the assumption that $z_1J_1=z_2J_2$, under which $\bar{\theta}_j = {\rm arctan}\frac{\Gamma}{H}\equiv \bar{\theta}$.
In a $D$-dimensional hypercubic lattice at $D\geq2$ ($D=2$ for a square lattice), seeking a plane wave solution for the fluctuations,
$\delta\theta_j=\theta_{\bm{p}}e^{i\left(\bm{p}\cdot\bm{x}_j-\omega t\right)}$, $\delta\varphi_j=\varphi_{\bm{p}}e^{i\left(\bm{p}\cdot\bm{x}_j-\omega t\right)}$, one obtains the dispersion relation
\begin{equation}
\begin{split}
    \omega=&\biggr[4\Gamma\biggr\{\Gamma+H\frac{\cos{\bar{\theta}}}{\sin{\bar{\theta}}}-\frac{z_1J_1}{\sin{\bar{\theta}}}+z_2J_2\left(\frac{\mathrm{cos^2}\bar{\theta}}{\sin{\bar{\theta}}}-\sin{\bar{\theta}}\right)\\
    &\!\!\!\! +4J_1\sin{\bar{\theta}}\sum_\alpha\mathrm{sin^2}\left(\frac{p_{\alpha}a-\pi}{2}\right)+4J_2\sin{\bar{\theta}}\\
   &\!\!\!\! \times\sum_{\alpha<\beta}\left(\mathrm{sin^2}\left(\frac{p_{\alpha}a+p_{\beta}a}{2}\right)\!+\!\mathrm{sin^2}\left(\frac{p_{\alpha}a-p_{\beta}a}{2}\right)\right)\biggr\}\biggr]^{\frac{1}{2}},
   \label{eq:dispersionMF}
\end{split}
\end{equation}
where $p_\alpha$ denotes the $\alpha$th component of the $D$-dimensional vector ${\bm p}$.
In a long-wavelength region, namely $|p_\alpha a -\pi|\ll1$, Eq.~(\ref{eq:dispersionMF}) can be approximated as
\begin{multline}
    \omega\simeq \biggr[4\Gamma\biggl\{ \Gamma+H\frac{\cos{\bar \theta}}{\sin{\bar \theta}}-\frac{z_1J_1}{\sin{\bar \theta}}+z_2J_2\left(\frac{\mathrm{cos^2}\bar{\theta}}{\sin{\bar \theta}}-\sin{\bar \theta}\right)\\
    +4\Gamma\sin{\bar \theta}\left(J_1+2J_2\left(D-1\right)\right)\sum_{\alpha}\left(p_\alpha a-\pi\right)^2\biggr\}\biggr]^\frac{1}{2}.
    \label{eq:bunsankankei_expand}
\end{multline}
Comparing this dispersion relation Eq.~(\ref{eq:bunsankankei_expand}) with that obtained by means of the GL theory Eq.~(\ref{eq:bunsankankei_GL}), the constants $w$ and $C$ in the GL equation are determined as
\begin{gather}
    w=\frac{s}{4\Gamma\left(\Gamma+H\cot{\bar{\theta}}-\frac{z_1J_1}{\sin{\bar{\theta}}}+z_2J_2\left(\frac{\mathrm{cos}^2\bar{\theta}}{\sin{\bar \theta}}-\sin{\bar \theta}\right)\right)},\label{eq:w}\\
    C= w\times4\Gamma\sin{\bar{\theta}}(J_1+2J_2(D-1))a^2.
\end{gather}
Thus, all the coefficients in the GL equation Eq.~(\ref{eq:GL}) have been explicitly related to the parameters in the original Ising model of Eq.~(\ref{eq:HamiltonianNNN}).

\begin{figure}[tbp]
\includegraphics[width=\linewidth]{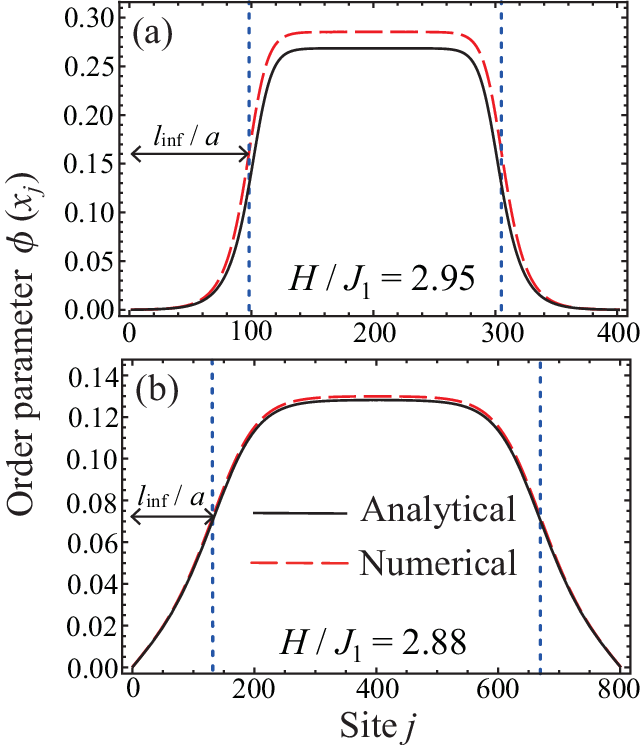}
\caption{Spatial profile of the antiferromagnetic order parameter for $(\Delta\Gamma/J_1,H/J_1)=(2.7\times10^{-8},2.95)$ (a) and $(4.1\times10^{-5},2.88)$ (b), where $\Delta\Gamma = \Gamma_{\rm t} - \Gamma$ and $\Gamma_{\rm t}$ corresponds to the transition point for a given value of $H/J_1$. The values of $\Gamma_{\rm t}/J_1$ within the GL theory are 5.456029819 for $H/J_1=2.95$ and 5.670719614 for $H/J_1=2.88$. Those determined by numerically solving the mean-field equations (\ref{eq:stationaryMF1}) and (\ref{eq:stationaryMF2}) for an infinite homogeneous system are 5.458428663 for $H/J_1=2.95$ and 5.670741170 for $H/J_1=2.88$. The black solid lines represent the analytical solution Eq.~(\ref{eq:phiGL}) while the red dashed lines represent the numerical solution of the mean-field equation (\ref{eq:stationaryMF1}) and (\ref{eq:stationaryMF2}). The blue dotted line marks the location of the inflection point of $\phi(x)$ for the numerical solution.}
\label{fig:mj}
\end{figure}

\subsection{Comparison between analytical and numerical analyses}
On the basis of an analytical solution of the stationary GL equation (\ref{eq:stationaryGL}) in the presence of a hard wall potential, we briefly review the surface criticality that emerges near the first-order quantum phase transition~\cite{Lipowsky,Lipowsky87,Danshita}. We focus on the case of the antiferromagnetic phase at two spatial dimensions and assume that the system is homogeneous in the $y$ direction. We also assume that there exists a hard wall potential at $x<0$, which leads to the boundary condition that $\phi(x=0)=0$. At a distance sufficiently far from the boundary, the order parameter is approximately uniform as 
\begin{eqnarray}
\phi(x)\simeq \phi_0 = \sqrt{\frac{-u+\sqrt{u^2-4sv}}{2v}}.
\end{eqnarray}
In Fig.~\ref{fig:phaseD}, the dotted line represents the contour of $\phi_0=0.25$ so that the gray-shaded region means the region of the antiferromagnetic phase in which  $\phi_0<0.25$. Since the condition $\phi(x)\ll 1$ is required for the GL approximation to be justified, the gray-shaded region roughly estimates the validity region of the GL theory.
For the transition to be of first order, the condition that $u<0$ has to be satisfied. In this case, the antiferromagnetic state is the ground state when $s<\frac{3u^2}{16v}$. These conditions imply that $\frac{u}{v\phi_0^2}=-\frac{4}{3}$ at the first-order transition point. Under the above-mentioned boundary conditions, one can solve Eq.~(\ref{eq:stationaryGL}) to obtain
\begin{equation}
    \phi(x)=\frac{\phi_0\sqrt{\gamma} \tanh{\frac{x}{\xi}}}{\sqrt{1+\gamma-\mathrm{tanh^2}\frac{x}{\xi}}},
    \label{eq:boundaryphi}
\end{equation}
where $\xi = \sqrt{\frac{2C}{u\phi_0^2 + 2v\phi_0^4}}$ and $\gamma = 2+\frac{3u}{2v\phi_0^2}$ ($\gamma=0$ at the first-order transition). For a fixed value of $H/J_1$, $\gamma \propto \Delta \Gamma/J_1 \equiv (\Gamma_{\rm t}-\Gamma)/J_1$ near the first order transition point $\Gamma = \Gamma_{\rm t}$.

When $0<\gamma<\frac{1}{2}$, $\phi(x)$ of Eq.~(\ref{eq:boundaryphi}) has an inflection point at $x=l_{\rm inf}$, where $l_{\rm inf}$ is given by
\begin{eqnarray}
\frac{l_{\rm inf}}{\xi} =  2\,{\rm arctanh}(1-2\gamma)^{\frac{1}{2}}.
\label{eq:linf}
\end{eqnarray}
$l_{\rm inf}$ can be interpreted as a healing length of the order parameter in the sense that the order parameter that is zero at the boundary recovers up to $\sim \frac{\phi_0}{2}$ at $x=l_{\rm inf}$.
Near the first-order transition point, where $\gamma\ll 1$, $\frac{l_{\rm inf}}{\xi} \simeq -\ln\gamma$. This logarithmic divergence of the length scale $l_{\rm inf}$ is a universal characteristic of the surface criticality. 

Another important critical behavior 
appears in the order parameter amplitude at a certain location near the boundary. For concreteness, at $x=\xi$, the order parameter behaves as
\begin{eqnarray}
\frac{\phi(x=\xi)}{\phi_0} = \frac{\sqrt{\gamma}\tanh{1}}{\sqrt{1+\gamma-\tanh^2{1}}}\simeq \frac{\sqrt{\gamma}\tanh{1}}{\sqrt{1-\tanh^2{1}}}.
\end{eqnarray}
More generically, the order parameter is proportional to $\gamma^{\frac{1}{2}}$ when $x\sim \xi$.
\begin{figure}[tbp]
\includegraphics[width=\linewidth]{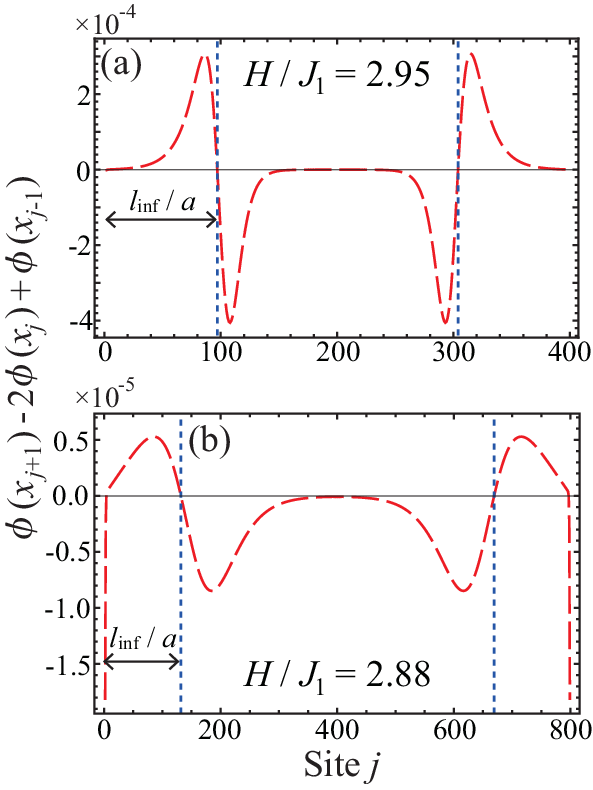}
\caption{Spatial profile of the approximate second derivative of the order parameter $(\phi(x_j + a) - 2\phi(x_j) + \phi(x_j - a))/a^2$ calculated by numerically solving the mean-field equations (\ref{eq:stationaryMF1}) and (\ref{eq:stationaryMF2}), where we set $a=1$. We take the same values of the parameters as in Fig.~\ref{fig:mj}. The blue dotted line marks the location of the inflection point of $\phi(x)$ for the numerical solution.}
\label{fig:ddmj}
\end{figure}

In actual calculations shown below, we consider the situation in which there is another hard wall at $x>(L+1)a$ in addition to the one at $x<0$. When $l_{\rm inf}$ is sufficiently small compared to the system size, the spatial dependence of the order parameter can be well approximated as
\begin{eqnarray}
\!\!\!\!\!\!\!\! \phi(x)\!=\!\left\{\begin{array}{cc}
 \frac{\phi_0\sqrt{\gamma} \tanh{\frac{x}{\xi}}}{\sqrt{1+\gamma-\mathrm{tanh^2}\frac{x}{\xi}}}, &{\rm if}\,\,0<\frac{x}{a}<\frac{L+1}{2}
 \\
 \frac{\phi_0\sqrt{\gamma} \tanh{\frac{(L+1)a-x}{\xi}}}{\sqrt{1+\gamma-\mathrm{tanh^2}\frac{(L+1)a-x}{\xi}}}, &{\rm if}\,\, \frac{L+1}{2}<\frac{x}{a}<L+1
 \end{array}\right. \!\!\!.
 \label{eq:phiGL}
\end{eqnarray}
In Fig.~\ref{fig:mj}, this solution for the two different values of $(\Delta\Gamma/J_1,H/J_1)$ is depicted by the black solid lines, where $\Delta \Gamma = \Gamma_{\rm t} - \Gamma$ and $\Gamma_{\rm t}$ corresponds to the first-order transition point for a given value of $H/J_1$.

With the analytical insights obtained from the solution of the GL equation in mind, we numerically analyze the surface criticality by solving the mean-field equations (\ref{eq:stationaryMF1}) and (\ref{eq:stationaryMF2}). Such numerical calculations are advantageous over the analytical ones in the sense that the former is applicable even when the order parameter at the transition is relatively large so that the validity of the GL equation can not be fully justified.
In our numerical calculations, we assume a two-sublattice structure in the $y$ direction and the open boundaries at $x_j=a$ and $La$ in the $x$ direction. 

\begin{figure}[tbp]
\includegraphics[width=\linewidth]{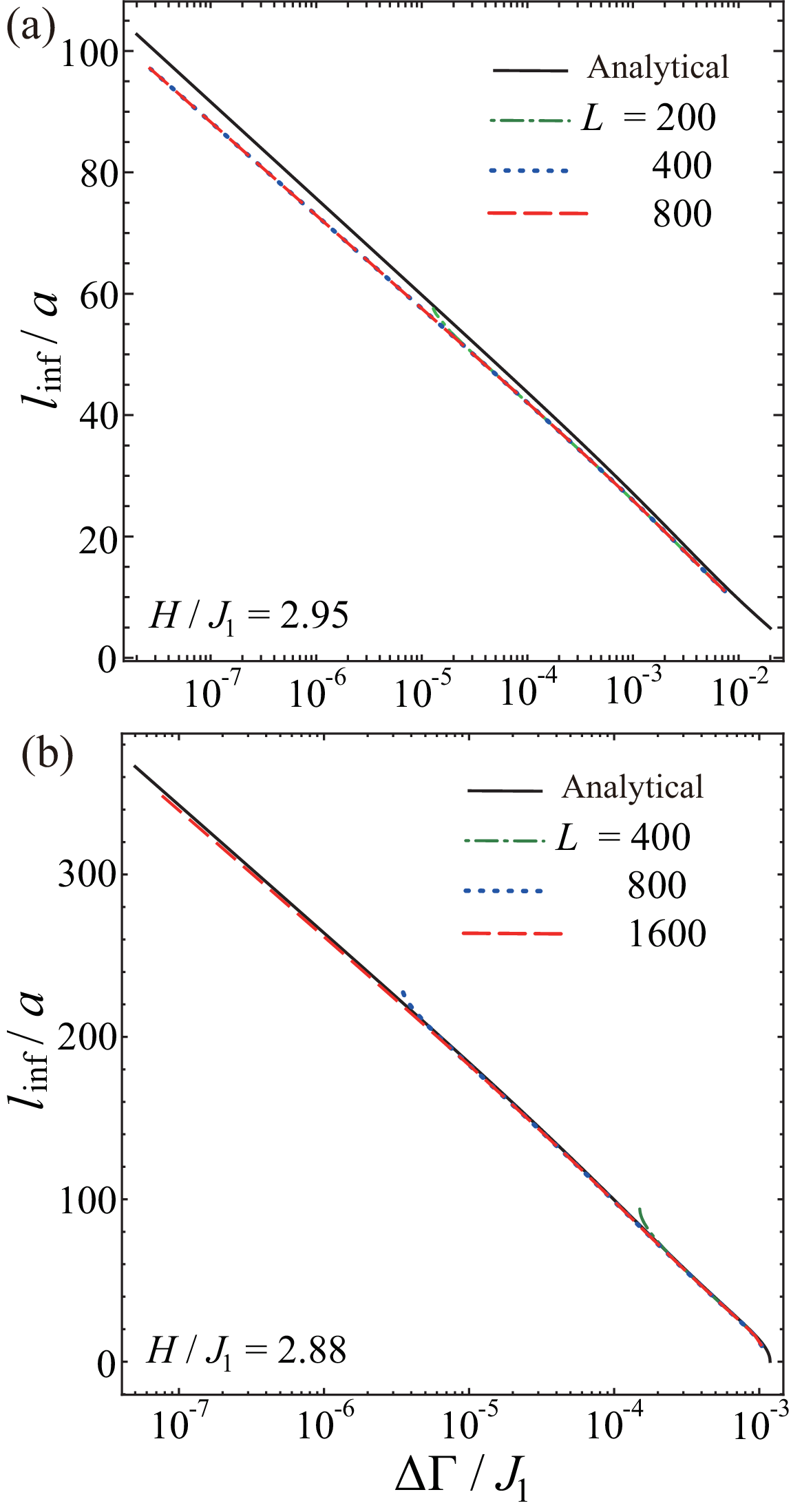}
\caption{Distance $l_{\rm inf}$ of the inflection point from the boundary as a function of $\Delta\Gamma/J_1$ for $H/J_1=2.95$ and 2.88. The black solid lines represent the analytical results of Eq.~(\ref{eq:linf})~\cite{Lipowsky,Lipowsky87,Danshita} while the dashed, dotted, and dash-dotted lines represent numerical results for different values of $L$. The longitudinal (horizontal) axis is in the linear (logarithmic) scale.}
\label{fig:linf}
\end{figure}

In Fig.~\ref{fig:mj}, we plot the spatial profile of the order parameters $\phi(x_j)$ computed from the mean-field equations for $(\Delta\Gamma/J_1,H/J_1)=(2.7\times10^{-8},2.95)$ (a) and $(4.1\times10^{-5},2.88)$ (b). In Fig.~\ref{fig:mj}(a), in which the parameter is relatively far from the QTCP, the deviation of the GL solution from the numerical solution is relatively large while the qualitative tendency of the spatial profile is well captured. This happens because the order parameter is too large, i.e., $\phi_0>0.25$, to justify the ignorance of the higher order terms in the GL expansion. By contrast, in Fig.~\ref{fig:mj}(b), since the system is closer to the QTCP, $\phi_0\simeq0.12$ such that the analytical result agrees quantitatively with the numerical one.

We next compare the distance to the inflection point from the boundary $l_{\rm inf}$. For this purpose, we calculate the second derivative of the order parameter by using the following approximate formula,
\begin{eqnarray}
\left.\frac{d^2\phi}{dx^2}\right|_{x=x_j} \simeq \frac{\phi(x_j + a) - 2 \phi(x_j) + \phi(x_j -a)}{a^2}.
\end{eqnarray}
In Fig.~\ref{fig:ddmj}, we show the spatial profile of the approximate second derivative of the order parameter for the same values of $(\Delta\Gamma/J_1,H/J_1)$ as in Fig.~\ref{fig:mj}. $l_{\rm inf}$ can be clearly determined as $x$ at which the second order derivative is equal to zero. 

In Fig.~\ref{fig:linf}, we show $l_{\rm inf}$ versus $\Delta\Gamma/J_1$ for different values of $L$. There we also plot the analytical expression of Eq.~(\ref{eq:linf}) for comparison. While in Fig.~\ref{fig:linf}(a) the agreement between the analytical and numerical results is only qualitative, in Fig.~\ref{fig:linf}(b) we see that the numerical results quantitatively agree with the analytical ones. Again, this happens because the parameters of the latter
are
rather close to the QTCP and $\phi_0$ is sufficiently small. Nevertheless, even in the former case, the numerically obtained $l_{\rm inf}$ exhibits clear logarithmic divergence. This manifests the universality of this surface critical behavior near the first-order transition point.

\section{Summary}
\label{sec:summary}
We have proposed how to realize the mixed-field Ising model with sign-inverted next-nearest-neighbor interaction in the system of Rydberg atoms in an optical tweezer array by using a Rydberg dressed Rydberg state. 
We have analyzed surface criticality, which is one of the interesting phenomena that can appear in the proposed model. 
We have derived a Ginzburg-Landau (GL) equation which describes the behavior of the antiferromagnetic order parameter near the quantum phase transition when the order parameter amplitude is small. All the GL coefficients have been explicitly related to the parameters in the original Ising model. 
We have also presented numerical calculations of the order parameter near the first-order transitions on the basis of a mean-field theory in order to directly compare them with the analytical insights based on the GL equation. We have found a quantitative agreement between the numerical and analytical results near the quantum tricritical point (QTCP).  In a parameter region away from QTCP, while the agreement is only qualitative, the healing length of the order parameter exhibits the logarithmic divergence that is a characteristic of the surface criticality, indicating the universality of this behavior.

\begin{acknowledgments}
The authors thank D.~Kagamihara, M.~Kunimi, and D.~Yamamoto for useful comments. This work was supported by JSPS KAKENHI (Grants No.~JP21H01014, No.~JP21K13855, and No.~JP24H00973), and MEXT Q-LEAP (Grant No.~JPMXS0118069021), JST FOREST (Grant No.~JPMJFR202T).
\end{acknowledgments}



\begin{thebibliography}{99}
\addcontentsline{toc}{section}{Reference}
\bibitem{Browaeys}A. Browaeys and T. Lahaye, Nat. Phys. \textbf{16}, 132 (2020).
\bibitem{Morgado} M. Morgado and S. Whitlock, AVS Quantum Sci. {\bf 3}, 023501 (2021).
\bibitem{Wu} X. Wu, X. Liang, Y. Tian, F. Yang, C. Chen, Y.-C. Liu, M. K. Tey, and L. You, Chin. Phys. B {\bf 30}, 020305 (2021).
\bibitem{Bernien} H. Bernien, S. Schwartz, A. Keesling, H. Levine, A. Omran, H. Pichler, S. Choi, A. S. Zibrov, M. Endres, M. Greiner, V. Vuleti\'c, and M. D. Lukin, Nature {\bf 551}, 579 (2017).
\bibitem{Ebadi} S. Ebadi, T. T. Wang, H. Levine, A. Keesling, G. Semeghini, A. Omran, D. Bluvstein, R. Samajdar, H. Pichler, W. W. Ho, S. Choi, S. Sachdev, M. Greiner, V. Vuletic, and M. D. Lukin, Nature \textbf{595}, 227 (2021).
\bibitem{Scholl} P. Scholl, M. Schuler, H. J. Williams, A. A. Eberharter, D. Barredo, K.-N. Schymik, V. Lienhard, L.-P. Henry, T. C. Lang, T. Lahaye, A. M. L\"auchli, and A. Browaeys, Nature {\bf 595}, 233 (2021).
\bibitem{Lienhard} V. Lienhard, S. de L\'es\'eleuc, D. Barredo, T. Lahaye, A. Browaeys, M. Schuler, L.-P. Henry, and A. M. L\"auchli, Phys. Rev. X {\bf 8}, 021070 (2018).
\bibitem{Rader} M. Rader and M. L\"auchli, arXiv:1908.02068 [cond-mat.quant-gas].
\bibitem{Kaneko} R. Kaneko, Y. Douda, S. Goto, and I. Danshita, J. Phys. Soc. Jpn. {\bf 90}, 073001 (2021).
\bibitem{Henkel} N. Henkel, R. Nath, and T. Pohl, Phys. Rev. Lett. \textbf{104},  195302 (2010).
\bibitem{Zeiher} J. Zeiher, R. van Bijnen, Peter Schau\ss, S. Hild, J.-Y. Choi, T. Pohl, I. Bloch, and C. Gross, Nat. Phys. \textbf{12}, 1095 (2016).
\bibitem{Beccaria} M. Beccaria, M. Campostrini, and A. Feo, Phys. Rev. B \textbf{76}, 094410 (2007). 
\bibitem{Kato} Y. Kato and T. Misawa, Phys. Rev. B \textbf{92}, 174419 (2015).
\bibitem{Lipowsky} R. Lipowsky, Phys. Rev. Lett. \textbf{49}, 1575 (1982).
\bibitem{Lipowsky87} R. Lipowsky, Ferroelectrics {\bf 73}, 69 (1987).
\bibitem{Danshita} I. Danshita, D. Yamamoto, and Y. Kato, Phys. Rev. A \textbf{91}, 013630 (2015).
\bibitem{Beguin} L. B\'eguin, A. Vernier, R. Chicireanu, T. Lahaye, and A. Browaeys, Phys. Rev. Lett. {\bf 110}, 263201 (2013)
\bibitem{Frenken} J. W. M. Frenken and J. F. van der Veen, Phys. Rev. Lett. {\bf 54}, 134 (1985). 
\bibitem{Bluhm} H. Bluhm, D. F. Ogletree, C. S. Fadley, Z. Hussain, and M. Salmeron, J. Phys.: Condens. Matter {\bf 14}, L227 (2002). 
\bibitem{Limmer} D. T. Limmer and D. Chandler, J. Chem. Phys. {\bf 141}, 18C505 (2014). 
\bibitem{Samajdar} R. Samajdar, W. W. Ho, H. Pichler, M. D. Lukin, and S. Sachdev, Phys. Rev. Lett. {\bf 124}, 103601 (2020).
\bibitem{Kalinowski} M. Kalinowski, R. Samajdar, R. G. Melko, M. D. Lukin, S. Sachdev, and S. Choi, Phys. Rev. B {\bf 105}, 174417 (2022).
\bibitem{Lipowsky83} R. Lipowsky, D. M. Kroll, and R. K. P. Zia, Phys. Rev. B {\bf 27}, 4499 (1983). 
\bibitem{Gompper} G. Gompper and D. M. Kroll, Phys. Rev. B {\bf 38}, 459 (1988). 
\bibitem{Danshita-10} I. Danshita and D. Yamamoto, Phys. Rev. A {\bf 82}, 013645 (2010).
\end{thebibliography}

\onecolumngrid  

\end{document}